\documentclass[aps,prl,twocolumn,showpacs,preprintnumbers,amsmath,amssymb, superscriptaddress]{revtex4}

\usepackage{graphicx}
\usepackage{dcolumn}
\usepackage{bm}

\begin{document}

\preprint{APS/123-QED}

\title
{
Demonstration of an optical quantum controlled-NOT gate without path interference
}

\author{Ryo Okamoto}
\affiliation{%
Research Institute for Electronic Science, Hokkaido University,
Sapporo 060--0812, Japan
}%
\author{Holger F. Hofmann}
\affiliation{%
Graduate School of Advanced Sciences of Matter, Hiroshima University Hiroshima 739-8530, Japan
}%
\author{Shigeki Takeuchi}
\affiliation{%
Research Institute for Electronic Science, Hokkaido University,
Sapporo 060--0812, Japan
}%
\author{Keiji Sasaki}
\affiliation{%
Research Institute for Electronic Science, Hokkaido University,
Sapporo 060--0812, Japan
}%

\date{\today}
%

\begin{abstract}
We report the first experimental demonstration of an optical
quantum controlled-NOT gate without any path interference,
where the two interacting path interferometers of the original
proposals (Phys. Rev. A {\bf 66}, 024308 (2001),
Phys. Rev. A {\bf 65}, 012314 (2002)) have been replaced by
three partially polarizing beam splitters with suitable polarization
dependent transmittances and reflectances. The performance
of the device is evaluated using a recently proposed method
(Phys. Rev. Lett. {\bf 94}, 160504 (2005)), by which the quantum
process fidelity and the entanglement capability can be estimated
from the 32 measurement results of two classical truth tables,
significantly less than the 256 measurement results required
for full quantum tomography.
\end{abstract}

\maketitle
%
Quantum computing promises to solve problems such as factoring large integers \cite{SHOR97} and searching over a large database \cite{GRO97} efficiently. One of the greatest challenges is to implement the basic elements of quantum computation in a reliable physical system and to
evaluate the performance of the operation in a sufficient manner.
In one of the earliest proposals for implementing quantum computation\cite{MIL88}, each qubit was encoded in a single photon existing in two optical modes. The main advantage of the photonic implementation of qubits is the robustness against decoherence and the availability of one-qubit operations. However, the difficulty
of realizing the nonlinear interactions between photons that are
needed for the implementation of two-qubit operations has been a
major obstacle. In recent work, Knill, Laflamme and Milburn (KLM) \cite{KLM01} have shown that this obstacle can be overcome by using
linear optics, single photon sources and photon number detectors.
By now, various controlled-NOT (CNOT) gates for photonic qubits
using linear optics
have been proposed \cite{KYI01, Pit01, Hof01, Ral01, Ral02, Nie04}
and demonstrated \cite{ Pit02, San02, Pit03, Bir03, Bir04, Zhao05}.

In particular, it has been shown in \cite{Hof01, Ral01} that a
`compact' CNOT gate can be realized by interaction at a single
beam splitter and post-selection of the output. Since this gate
requires no ancillary photon inputs or additional detectors, it
should be especially useful  for experimental realizations of optical
quantum circuits.
However, there have been two crucial difficulties. In the original scheme\cite{Hof01, Ral01}, the polarization sensitivity of the operation was achieved by separating the paths of the orthogonal polarizations, essentially creating two interacting two-path interferometers. Therefore, the initial experimental realizations based the original proposal \cite{Bir03,Bir04} are very sensitive to the noisy environment
(thermal drifts and vibrations), making it necessary to control and
to stabilize nanometer order path-length differences. In addition to
these problems, perfect mode-matching is required in each output of the interferometer. Thus, it is very difficult to construct quantum circuits using devices based on those experimental setups. Another difficulty is the evaluation of experimental errors in multi qubit gates.
In order to obtain the most complete evaluation of gate performance
possible, quantum process tomography has been used in the previous experiment \cite{Bir04}. However, 256 different measurement setups are required to evaluate only one CNOT device. When we have to evaluate
even more complicated quantum devices realized by a combination of
gates, the number of measurements required for tomography rapidly
increases as the number of input and output qubits increases.

In this paper, we present an experimental realization of the
`compact' optical CNOT gate \cite{Hof01, Ral01} without any
path interference. We show that the CNOT gate can be
implemented using three partially polarizing beam splitters (PPBSs)
with suitable polarization dependent transmittances and reflectances,
where the essential interaction is realized by a single intrinsic
PPBS, while the other two supplemental PPBSs act as local polarization compensators on the
input qubits. The gate operation can then be obtained directly
from the polarization dependence of the reflectances of the PPBSs,
removing the need for interference between different paths for
orthogonal polarizations.
We have evaluated the device operation using a recently proposed
method \cite{Hof04}, by which we can determine the lower and upper
bounds of the process fidelity from measurements of only 32
input-output combinations. We can thus characterize the gate operation with 1/8 the number
of input-output measurements required for complete quantum process tomography.
We hope that these results will open a door to the realization of
more complex quantum circuits for quantum computing.


Figure 1(a) shows the previously proposed optical circuit for the CNOT gate \cite{Hof01, Ral01}.
The beam splitter sitting in the center of the circuit is the essential one which realizes the quantum phase gate operation by flipping the
phase of the state $|H; H \rangle$, where both photons are horizontally
polarized, to $-|H; H \rangle$ due to two-photon interference.
Since this operation attenuates the amplitudes of horizontally
polarized components by a factor of $1/\sqrt{3}$, the other
two beam splitters with reflectivity 1/3 are inserted in each of the interferometer paths in order to also attenuate the amplitudes of vertically polarized components $|V\rangle$, so that the total
amplitude of any two photon input is uniformly attenuated to $1/3$.
If we now define the computational basis of the gate as
$|0_z\rangle _{C} \equiv |V\rangle _{C}$, $|1_z\rangle _{C} \equiv |H\rangle _{C}$ for the control qubit, and $|0_z\rangle _{T}\equiv 1/\sqrt{2} (|V\rangle _{T}+|H\rangle _{T})$, $|1_z\rangle _{T} \equiv 1/\sqrt{2} (|V\rangle _{T} -|H\rangle _{T})$ for the target qubit, the gate performs the unitary operation
$\hat{U}_{\mbox{\small CNOT}}$ of the quantum CNOT on the input qubits.

The difficulty in the original proposal is that we have to stabilize the two interferometers
of the horizontally and vertically polarized paths by controlling the length of four optical paths with an accuracy on the order of nanometers in order to achieve a reliable operation of the device. In addition, the modes in each path have to be aligned precisely at the output ports
of the polarizing beam splitters. Such difficulties have been crucial obstacles for the future realization of optical quantum circuits consisting of several CNOT gates.

%

\begin{figure}[ht]
\scalebox{0.35}[0.35]{
\includegraphics{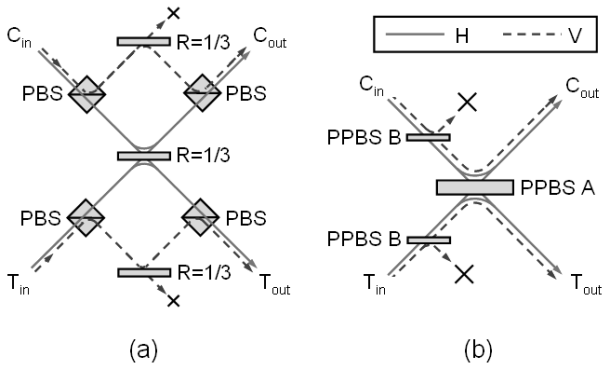}}
\caption{\label{fig:CNOT} Schematics of the `compact' CNOT gate.
(a) The optical circuit in the original proposals \cite{Hof01, Ral01}. Here, the polarizing beam splitters (PBSs) reflect ( transmit ) photons with vertical ( horizontal ) polarization. (b) The optical circuit
without any path interference using partially polarizing beam splitters.}
\end{figure}

Fig. 1(b) shows our solution to this problem. We use one intrinsic PPBS (PPBS-A) and two supplemental PPBSs (PPBS-B) in the optical circuit.
The intrinsic PPBS-A, which corresponds to the central beam splitter in Fig. 1(a), implements the quantum phase gate operation by
reflecting vertically polarized light perfectly and reflecting
(transmitting) 1/3 (2/3) of horizontally polarized light.
The two supplemental PPBS-Bs are inserted to adjust the
amplitudes of the local horizontal and vertical components of the
photonic qubits by transmitting (reflecting) 1/3 (2/3) of vertically
polarized light and transmitting horizontally polarized light perfectly.
As Fig. 1(b) shows, the use of PPBSs allows us to
reduce the four optical paths in Fig. 1(a) to only two optical paths,
and path-interferometers are no longer required for the implementation
of the `compact' quantum CNOT gate.

In the following experimental demonstration, we used a simple
polarization compensation instead of the two supplemental PPBS-Bs.
The only purpose of the PPBS-Bs is to reduce the amplitude of the
vertical component in the input to $1/\sqrt{3}$ of the original
input value, while leaving the horizontal component unchanged.
Therefore, we can easily simulate the function of the supplemental
PPBS-Bs by using compensated input states whose vertical
component is reduced to $ 1/\sqrt{3}$. To simulate a general input state
$|\psi_{\mbox{\small effective}} \rangle = c_H |H\rangle + c_V |V\rangle$, we thus  use a compensated input state of
$|\psi_{\mbox{\small comp.}}\rangle = c_H |H\rangle + (c_V/\sqrt{3}) |V\rangle$. For example, the input for the target qubit
state $|0_z\rangle_T$ becomes $|0_{z}^\prime\rangle_T = (\sqrt{3}|H\rangle+|V\rangle)/\sqrt{6}$, which can be easily prepared just
by changing the angle of linear polarization by rotating the $\lambda$/2 plate (HWP) in the target input.


The schematic of our experimental setup is shown in Fig.2. We used a pair of photons generated through spontaneous parametric down conversion for our input. A beta barium borate (BBO) crystal cut for the type-II twin-beam condition \cite{Tak01} was pumped by an argon ion laser at a wavelength of 351.1 nm. The pump beam was focused in the BBO crystal using a convex lens to increase the photon flux\cite{Kur01}. Pairs of twin photons are emitted at 702.2 nm with orthogonal polarizations. Gland-Thompson polarizers are used to increase the extinction ratio.
After removing the scattered pump light using bandpass filters (IF, center wavelength 702.2nm, FWMH 0.3nm), each of the photons was guided to polarization maintaining single-mode fibers (PMFs) via objective lens and then delivered to the CNOT verification setup.
 After the collimation lens for the output of the PMFs, the polarization of photons are controlled by HWPs mounted in rotating stages. The timing of the two entangled photons injected to the PPBS-A was controlled using an optical delay. A $\lambda/4$ plate (QWP) was inserted to compensate the phase change between horizontal and vertical polarization in the optical delay. After the quantum interference which occurs at PPBS-A, the polarization of output photons was analyzed using HWPs and PBSs. Finally, those photons are coupled into single mode fibers (SMFs) and counted by the single photon counters (SPCM-AQ-FC, Perkin Elmer).

\begin{figure}[ht]
\scalebox{0.35}[0.35]{
\includegraphics{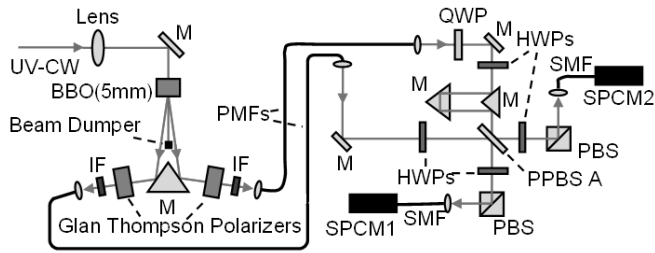}}
\caption{\label{fig:ExpSetup} Experimental setup for the demonstration of the CNOT gate without any path interference. M is for reflecting mirrors.}
\end{figure}
%
%
Our setup permits us to select various linear polarizations for the input, and to detect the corresponding linear polarizations
in the output. As has been shown in \cite{Hof04}, it is possible to characterize the essential quantum properties of the gate
operation by using the computational $ZZ$-basis given above and the complementary linearly polarized $XX$-basis given by
$|0_x\rangle _{C} \equiv  1/\sqrt{2} (|V\rangle _{C}+|H\rangle _{C})$,
$|1_x\rangle _{C} \equiv 1/\sqrt{2} (|V\rangle _{C} -|H\rangle _{C})$ for the control qubit,
and  $|0_x\rangle _{T}\equiv |V\rangle _{T}$, $|1_x\rangle _{T} \equiv |H\rangle _{T}$ for the target qubit.
The operation of the gate on this input basis also corresponds to a CNOT, with the target qubit acting
on the control qubit (reverse CNOT).

\begin{figure}[ht]
\scalebox{0.3}[0.3]{
\includegraphics{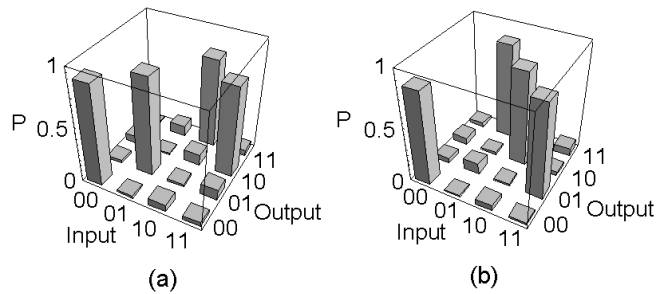}}
\caption{\label{fig:FzzFxx} (a) Bar graph of the experimental results for the CNOT operation in the $ZZ$ basis. (b) Bar graph of the experimental results for the reverse CNOT operation in the $XX$ basis.}
\end{figure}

The measurement result of the input-output probabilities of our CNOT gate in the $ZZ$ basis and in the $XX$ basis are shown in tables
\ref{data}(a) and \ref{data}(b), respectively. We measured the coincidence counts between two SPCMs by appropriately setting the HWPs for 16 different combinations of input and output states. In order to convert the coincidence rates to probabilities, we normalize them with the sum of coincidence counts obtained for the respective input state.
\begin{table}
\begin{ruledtabular}
\begin{tabular}{l|cccc}
(a) & \hspace{0.15cm} $\langle 0_z0_z|$ \hspace{0.15cm}
    & \hspace{0.15cm} $\langle 0_z1_z|$ \hspace{0.15cm}
    & \hspace{0.15cm} $\langle 1_z0_z|$ \hspace{0.15cm}
    & \hspace{0.15cm} $\langle 1_z1_z|$ \hspace{0.15cm}
\\ \hline
$|0_z0_z \rangle$ & 0.898 & 0.031 & 0.061 & 0.011
\\
$|0_z1_z \rangle$ & 0.021 & 0.885 & 0.006 & 0.088
\\
$|1_z0_z \rangle$ & 0.064 & 0.027 & 0.099 & 0.810
\\
$|1_z1_z \rangle$ & 0.031 & 0.096 & 0.819 & 0.054
\end{tabular}
\end{ruledtabular}

\vspace{0.2cm}

\begin{ruledtabular}
\begin{tabular}{l|cccc}
(b) & \hspace{0.15cm} $\langle 0_x0_x|$ \hspace{0.15cm}
    & \hspace{0.15cm} $\langle 0_x1_x|$ \hspace{0.15cm}
    & \hspace{0.15cm} $\langle 1_x0_x|$ \hspace{0.15cm}
    & \hspace{0.15cm} $\langle 1_x1_x|$ \hspace{0.15cm}
\\ \hline
$|0_x0_x \rangle$ & 0.854 & 0.044 & 0.063 & 0.039
\\
$|0_x1_x \rangle$ & 0.013 & 0.099 & 0.013 & 0.874
\\
$|1_x0_x \rangle$ & 0.050 & 0.021 & 0.871 & 0.058
\\
$|1_x1_x \rangle$ & 0.019 & 0.870 & 0.040 & 0.071
\end{tabular}
\end{ruledtabular}
\caption{\label{data}
Measurement results for the input-output probabilities of
the CNOT operation in the $ZZ$-basis (a) and the reverse
CNOT operation in the $XX$-basis (b).
}
\end{table}
The three dimensional bar graphs of table \ref{data} are shown in
Fig. \ref{fig:FzzFxx}. The fidelity $F_{zz}$ of the CNOT operation in the $ZZ$ basis, defined as the probability of
obtaining the correct output averaged over all four possible inputs, is $0.85$.
Similarly, the fidelity $F_{xx}$ of the reverse CNOT operation in the $XX$ basis is $0.87$.

As discussed in detail in \cite{Hof04}, the two complementary fidelities $F_{zz}$ and $F_{xx}$ define an upper and lower
bound for the quantum process fidelity $F_{\rm process}$ of the gate with
\begin{equation}
 F_{zz}+ F_{xx}-1 \le F_{\rm process} \le \min \{F_{zz}, F_{xx}\}
\end{equation}
Thus, our experimental results show that the process fidelity of our experimental quantum CNOT gate is
\begin{equation}
 0.72 \le F_{\rm process} \le 0.85 .
\end{equation}
The lower bound of the process
fidelity also defines a lower bound of the entanglement capability
of the gate, since the fidelity of entanglement generation
is at least equal to the process fidelity.
In terms of the concurrence $C$ that the gate can generate from
product state inputs, the minimal entanglement capability is
therefore given by \cite{Hof04}
\begin{equation}
 C \ge 2 F_{\mbox{\small process}} - 1.
\end{equation}
Since our experimental results show that the minimal process fidelity
of the gate is $0.72$, the lower bound of the entanglement capability
is
\begin{equation}
C \ge 0.44.
\end{equation}
The experimental results shown in table \ref{data} are therefore
sufficient to confirm the entanglement capability of our gate.


In order to gain a better understanding of the noise effects
in our experimental quantum gate, we can analyze the errors
in the classical operations shown in table \ref{data} and associate
them with quantum errors represented by elements of the
process matrix. For this purpose, it is useful to classify the
errors according to the bit flip errors in the output of the
operations in the $ZZ$ and the $XX$ basis, using 0 for the
correct gate operation, C for a flip of the control bit output,
T for a flip of the target
bit output, and B for a flip of both outputs.
It is then possible to expand the process matrix in terms of
16 orthogonal unitary operations $\hat{U}_i$, where
the index $i=\{$ 00, C0, B0, T0, 0C, CC, TC, BC, 0T, CT, TT,
BT, 0B, CB, TB, BB $\}$
defines the pair of error syndromes in the complementary operations in
the $ZZ$ and the $XX$ basis, e.g. $\hat{U}_{\mbox{TC}}$ for
a target flip error in the $ZZ$ operation and a control
flip error in the $XX$ operation and
$\hat{U}_{00}=\hat{U}_{\mbox{\small CNOT}}$ for the ideal
gate operation.
The process matrix describing the noisy gate operation is
given by the operator sum representation of the relation
between an arbitrary input density matrix
$\hat{\rho}_{\mbox{\small in}}$ and its output density
matrix $\hat{\rho}_{\mbox{\small out}}$,
\begin{equation}
\hat{\rho}_{\mbox{\small out}}
= \sum_{i,j} \chi_{i,j}\;\hat{U}_{i} \hat{\rho}_{\mbox{\small in}} \hat{U}_{j}^\dagger.
\end{equation}
Since each operation $\hat{U}_{i}$ describes a well defined
combination of errors in the $ZZ$ and $XX$ operations,
it is now possible to relate the error probabilities observed
in table \ref{data} to sums over the diagonal elements $\chi_{i,i}$
of the process matrix, as shown in table \ref{sums}.
\begin{table}

\begin{ruledtabular}
\begin{tabular}{c|cccc|c}
 $\chi_{i,i}$ & X0 & XC & XT & XB & sum \\
\hline
 \hspace{0.35mm}0\hspace{0.35mm}X & $\chi_{\mbox{00,00}}$ & $\chi_{\mbox{0C,0C}}$
    & $\chi_{\mbox{0T,0T}}$ & $\chi_{\mbox{0B,0B}}$ & $0.853$ \\
 CX & $\chi_{\mbox{C0,C0}}$ & $\chi_{\mbox{CC,CC}}$
    & $\chi_{\mbox{CT,CT}}$ & $\chi_{\mbox{CB,CB}}$ & $0.052$ \\
 TX & $\chi_{\mbox{T0,T0}}$ & $\chi_{\mbox{TC,TC}}$
    & $\chi_{\mbox{TT,TT}}$ & $\chi_{\mbox{TB,TB}}$ & $0.051$ \\
 BX & $\chi_{\mbox{B0,B0}}$ & $\chi_{\mbox{BC,BC}}$
    & $\chi_{\mbox{BT,BT}}$ & $\chi_{\mbox{BB,BB}}$ & $0.044$ \\
\hline
sum & $0.867$        & $0.071$        & $0.034$        & $0.028$        & $1.000$
\end{tabular}
\end{ruledtabular}
\caption{\label{sums}
Relation between experimentally observed errors and
process matrix elements. (X=0,C,T,B).
}
\end{table}
Since the correlations between errors in $ZZ$ and errors in $XX$
are unknown, a range of different distributions of diagonal
elements $\chi_{i,i}$ is consistent with the experimental data.
It is possible to illustrate this range of possibilities
by considering the scenarios of lowest and highest process fidelity.
The matrix elements of these cases are shown in tables \ref{pmat}(a) and
\ref{pmat}(b).
\begin{table}

\begin{ruledtabular}
\begin{tabular}{c|cccc|c}
 \rm{(a)} & X0 & XC & XT & XB & sum \\
\hline
 \hspace{0.35mm}0\hspace{0.35mm}X & $0.720$ & $0.071$ & $0.034$ & $0.028$ & $0.853$ \\
 CX & $0.052$ & $0.000$ & $0.000$ & $0.000$ & $0.052$ \\
 TX & $0.051$ & $0.000$ & $0.000$ & $0.000$ & $0.051$ \\
 BX & $0.044$ & $0.000$ & $0.000$ & $0.000$ & $0.044$ \\
\hline
sum & $0.867$ & $0.071$ & $0.034$ & $0.028$ & $1.000$
\end{tabular}
\end{ruledtabular}
\vspace{0.2cm}
\begin{ruledtabular}
\begin{tabular}{c|cccc|c}
\rm{(b)} & X0 & XC & XT & XB & sum \\
\hline
 \hspace{0.35mm}0\hspace{0.35mm}X & $0.853$ & $0.000$ & $0.000$ & $0.000$ & $0.853$ \\
 CX & $0.005$ & $0.025$ & $0.012$ & $0.010$ & $0.052$ \\
 TX & $0.005$ & $0.024$ & $0.012$ & $0.010$ & $0.051$ \\
 BX & $0.004$ & $0.022$ & $0.010$ & $0.008$ & $0.044$ \\
\hline
sum & $0.867$ & $0.071$ & $0.034$ & $0.028$ & $1.000$
\end{tabular}
\end{ruledtabular}

\caption{\label{pmat}
Diagonal elements of the process matrix for (a) the worst case
of process fidelity $0.72$ and (b) the optimal case of process
fidelity $0.85$.
}
\end{table}
In the worst case scenario of a process fidelity of $0.72$
shown in table \ref{pmat}(a),
each error syndrome is only observed in either the $ZZ$ or the
$XX$ basis. Therefore, the errors observed in the $ZZ$ operation
can be identified directly with $i=$C0,T0,B0,
and the errors observed in the $XX$ operation can be identified
with $i=$0C,0T,0B. In the optimal case of a process fidelity of
$0.85$ shown in table \ref{pmat}(b), the diagonal elements of
$i=$C0,T0,B0 for errors only in $ZZ$ are all zero. The remaining
errors have been distributed over the other diagonal matrix
elements within the constraints given by table \ref{sums}. Note that
the distribution of matrix elements
in the optimal case is far more homogeneous than the worst case
scenario. The actual process fidelity is therefore likely to be
closer to the upper limit of $0.85$ than to the lower limit
of $0.72$.

In conclusion, we have demonstrated the first experimental realization
of an optical quantum CNOT gate without any path interference by measuring the
fidelities of the classical CNOT operations in the computational
$ZZ$ basis and in the complementary $XX$ basis. The performance of
both operations by the same quantum gate at fidelities of $0.85$
and $0.87$ indicates that our device has a quantum process fidelity
of $0.72 \le F_{\mbox{\small process}} \le 0.85$ and an entanglement
capability of $C \ge 0.44$.
Since the gate presented in this paper requires no path length
adjustments, it should be ideal for the construction of
quantum circuits using multiple gates. The present work may
therefore provide an important first step towards the realization
of optical quantum computation with larger numbers of qubits.

%
%
We would like to thank H. Fujiwara, K. Tsujino and D. Kawase for technical suggestions.
This work was supported in part by 
Core Research for Evolutional Science and Technology, Japan Science and Technology Agency,
Grant-in-Aid of Japan Science Promotion Society and 21st century COE program.
%
%

%
\end{document}